\documentclass[aps,prb,twocolumn,notitlepage,showpacs,superscriptaddress,am]{revtex4-2}%
\usepackage{graphicx}
\usepackage{amsmath}
\usepackage{amssymb}
\usepackage{mhchem}
\usepackage{color}
\usepackage{multirow}
\usepackage{amsfonts}%
\usepackage[breaklinks=true,colorlinks=true,linkcolor=blue,urlcolor=blue,citecolor=blue]{hyperref}

\providecommand{\U}[1]{\protect\rule{.1in}{.1in}}

\begin{document}
\title{Putative quantum critical point in locally noncentrosymmetric \ce{CeCoGe2} crystals}

\author{F. Garmroudi}
\affiliation{Materials Physics Applications--Quantum, Los Alamos National Laboratory, Los Alamos, New Mexico 87545, USA}
\author{C.\,S.\,T. Kengle}
\affiliation{Materials Physics Applications--Quantum, Los Alamos National Laboratory, Los Alamos, New Mexico 87545, USA}
\author{M.\,H. Schenck}
\affiliation{Materials Physics Applications--Quantum, Los Alamos National Laboratory, Los Alamos, New Mexico 87545, USA}
\affiliation{Department of Physics, Colorado State University, Fort Collins, Colorado 80523, USA}
\author{J.\,D. Thompson}
\affiliation{Materials Physics Applications--Quantum, Los Alamos National Laboratory, Los Alamos, New Mexico 87545, USA}
\author{E.\,D. Bauer}
\affiliation{Materials Physics Applications--Quantum, Los Alamos National Laboratory, Los Alamos, New Mexico 87545, USA}
\author{S.\,M. Thomas}
\affiliation{Materials Physics Applications--Quantum, Los Alamos National Laboratory, Los Alamos, New Mexico 87545, USA}
\author{P.\,F.\,S. Rosa}
\affiliation{Materials Physics Applications--Quantum, Los Alamos National Laboratory, Los Alamos, New Mexico 87545, USA}


\begin{abstract}
Locally noncentrosymmetric heavy-fermion compounds may produce long-sought correlated quantum phases, such as spin-triplet superconductivity with non-Abelian quasiparticles, but identifying the right candidate systems is challenging. Here, using the In flux method, we synthesize \ce{CeCoGe2} single crystals, belonging to the highly tunable pseudotetragonal ($Cmcm$) \ce{Ce$TX$2} family, which allows for substitutions at both the transition metal $T$ and at the $X$ sites. We identify a heavy-fermion ground state with a Sommerfeld coefficient $\gamma\approx 120\,$mJ\,mol$^{-1}$\,K$^{-2}$ 
and a non-Fermi-liquid exponent of the electrical resistivity,
which may indicate its proximity to the putative quantum critical point. 
However, no signs of superconductivity or magnetic order are detected down to 20\,mK. Our analysis of electrical transport and structural properties indicates that coherent charge transport and the emergence of superconductivity observed under hydrostatic pressure in related compounds (\ce{CePtSi2} and \ce{CeRhGe2}) are suppressed in \ce{CeCoGe2} by strong random potential scattering due to intrinsic Co vacancies (approximately 4\,\% even in the highest-quality crystals). By tuning the growth stoichiometry and temperature profile, we demonstrate that the defect concentration can be controlled and has a pronounced effect on the residual resistivity. We hypothesize that superconductivity may be found in higher-quality \ce{CeCoGe2} crystals grown by different techniques.
\end{abstract}

\maketitle

\section{introduction}
The study of 4$f$ and 5$f$ electron-based heavy-fermion compounds has long been a fertile ground for exploring the rich interplay between strong electronic correlations, magnetism, and the exotic quantum phases emerging therefrom \cite{stewart1984heavy,fisk1988heavy,park2006hidden,monthoux2007superconductivity,si2010heavy,mydosh2011colloquium}. One of the most intriguing phenomena emerging from this interplay is spin-triplet superconductivity -- an unconventional and exceptionally rare superconducting state in which Cooper pairs form odd-parity orbital wavefunctions, in contrast to the spin-singlet, even-parity pairing in conventional superconductors \cite{sigrist1991phenomenological,mackenzie2003superconductivity,huxley2001uge,joynt2002superconducting,ran2019nearly}. Beyond its fundamental interest, such a superconducting state can exhibit nontrivial topological properties, giving rise to exotic quasiparticles such as Majorana zero modes, which are considered promising platforms for realizing topologically protected, fault-tolerant quantum computation \cite{nayak2008non,sarma2015majorana,flensberg2021engineered}.
In particular, the realization of a chiral, multicomponent $p_x+ip_y$ pairing state has been a long-standing goal \cite{read2000paired,volovik2003universe,rice1995sr2ruo4}. In two dimensions, such a pairing symmetry would be fully gapped, allowing the Majorana zero modes to be protected and energetically separated from other low-energy excitations.

Key structural ingredients to realize such an exotic superconductor in a bulk heavy-fermion material appear to be (i) local inversion symmetry breaking in a globally centrosymmetric crystal structure; more specifically, there should exist two or more $f$-electron ions per unit cell related in pairs by inversion \cite{anderson1985further,hazra2023triplet} and (ii) a two-dimensional, layered crystal structure, where superconductivity in the layers is sufficiently immune to interlayer coupling. Among all known spin-triplet superconductor candidates, the majority are $f$-electron materials, with only a handful of U-based compounds (\ce{UBe13} \cite{walti2000spectroscopic,shimizu2019spin}, \ce{UPt3} \cite{tou1998nonunitary}, U(Co,Rh)Ge \cite{hattori2012superconductivity,aoki2019review}, \ce{U2PtC2} \cite{mounce2015detection} and \ce{UTe2} \cite{ran2019nearly,aoki2022unconventional}) and only a single Ce-based system (\ce{CeRh2As2} \cite{khim2021field}) displaying signatures of spin-triplet superconductivity at ambient pressure. Additionally, a few other systems have been identified under pressure \cite{saxena2000superconductivity,o2022multicomponent,li2024unconventional,squire2023superconductivity,shi2026superconductivity}. However, none of these materials has been shown to realize the sought-after $p_x+ip_y$ pairing state. It is thus crucial to identify and systematically screen other candidates.

\begin{figure}[t!]
\centering
 \includegraphics[width=0.45\textwidth]{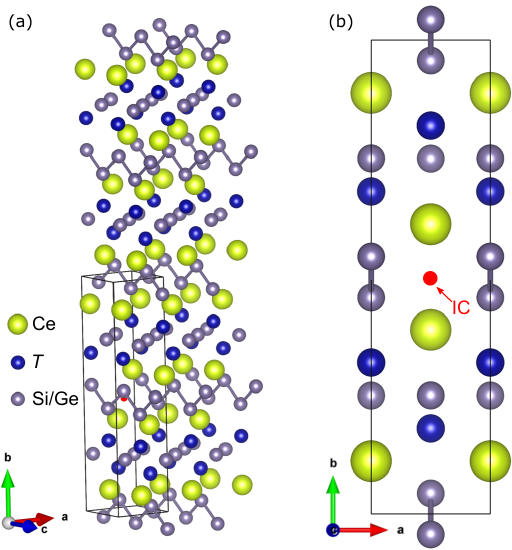}
\caption{(a) Pseudotetragonal crystal structure ($a \approx c \ll b$) of \ce{Ce$TX$2} ($T$ being a transition metal and $X=$ Si, Ge) with the locally noncentrosymmetric \ce{CeNiSi2} structure ($Cmcm$). (b) Ce atoms are positioned around an inversion center (IC) located in the middle of the unit cell. 
}
\label{Fig1}
\end{figure} 

\begin{figure}[t!]
\centering
 \includegraphics[width=0.42\textwidth]{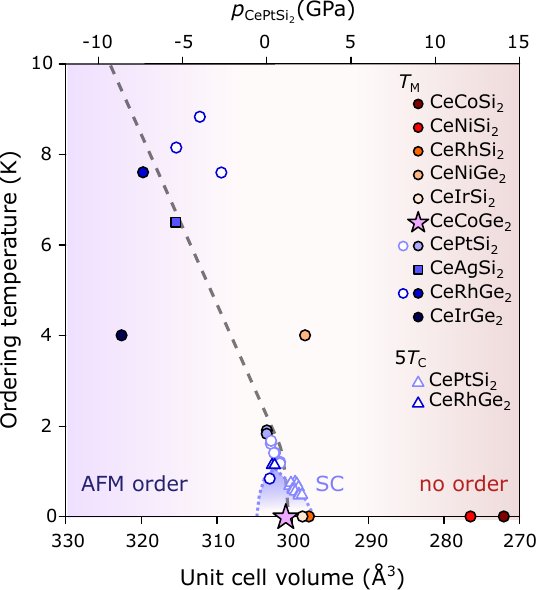}
\caption{Phase diagram and putative quantum critical point in locally noncentrosymmetric \ce{Ce$TX$2}. Chemical (filled symbols) and hydrostatic (open symbols) pressure tune the magnetic ground state from antiferromagnetic order for a unit-cell volume $V\gtrsim 300$\,\AA$^3$ to paramagnetic states for $V\lesssim 300$\,\AA$^3$. A superconducting dome emerges for \ce{CePtSi2} and \ce{CeRhGe2} under hydrostatic pressure, centered around $V_\text{c}\approx 300$\,\AA$^3$, which corresponds to the unit-cell volume of \ce{CeCoGe2}.
}
\label{Fig2}
\end{figure} 

Here, we investigate \ce{CeCoGe2} belonging to the \ce{Ce$TX$2} family ($T$ being a transition metal and $X=$ Si, Ge), which crystallize in the pseudotetragonal, orthorhombic ($Cmcm$) \ce{CeNiSi2} structure [Fig.\,1(a)], which has an inversion center in the center of the unit cell with Ce atoms occupying sites that lack local inversion symmetry [see Fig.\,1(b)] \cite{francois1990nouveaux,lee1990crystal}. The hybridization between the localized Ce $4f$ states and the conduction electrons in these systems, is highly sensitive to the unit-cell volume ($V$), giving rise to antiferromagnetically-ordered states for $V\gtrsim 300\,${\AA}$^3$ \cite{lee1990crystal,cordruwisch2001constitution,jung2002magnetocrystalline,hirose2011magnetic,szlawska2018antiferromagnetic} and paramagnetic ground states if the unit-cell volume is less than $\approx300\,${\AA}$^3$ (see Fig.\,2) \cite{mun2003anderson,pelizzone1982magnetic,adroja1993structural,chevalier1993intermediate}. In addition, the period of the transition metal strongly affects the $c$--$f$ hybridization, with $3d$ transition metals favoring more localized, magnetic behavior, whereas the $5d$ elements tend to promote delocalization (compare for instance \ce{CeNiGe2} and \ce{CeIrGe2} in Fig.\,2). It has been shown that both \ce{CePtSi2} and \ce{CeRhGe2} become superconducting when their unit-cell volume is tuned via hydrostatic pressure to a putative quantum critical point (QCP) \cite{nakano2009pressure,hirose2011magnetic}. Using pressure-dependent x-ray diffraction data for \ce{CePtSi2} from Ref.\,\cite{oomi1994effect}, we convert hydrostatic pressure to unit-cell volume and find that the critical pressures, $p_\text{c}\approx 1.2\,$GPa for \ce{CePtSi2} \cite{nakano2009pressure} and $p_\text{c}\approx 7\,$GPa for \ce{CeRhGe2} \cite{hirose2011magnetic}, correspond to a common critical volume of $V_\text{c}\approx 300-301\,${\AA}$^3$. Nakano \textit{et al.} first demonstrated pressure-induced superconductivity in polycrystalline \ce{CePtSi2} \cite{nakano2009pressure}, and Hirose \textit{et al.} subsequently extended their study on single crystals \cite{hirose2011magnetic}. We note that, although not explicitly discussed in Ref.\,\cite{hirose2011magnetic}, their findings suggest strongly enhanced upper critical fields that exceed the Pauli limiting field by nearly an order of magnitude --  a hallmark of spin-triplet superconductivity. Together with the essential symmetry ingredients in the crystal structure, these observations make a case for \ce{CePtSi2} and \ce{CeRhGe2} as candidate spin-triplet superconductors under pressure. 

Plotting the magnetic and superconducting transition temperatures of various \ce{Ce$TX$2} systems versus their unit-cell volume (Fig.\,2) further suggests that among all these compounds, \ce{CeCoGe2} is positioned closest to the putative QCP and the associated superconducting dome. However, to date, no experimental studies below 2\,K have been reported on \ce{CeCoGe2}, and superconductivity in \ce{CePtSi2} and \ce{CeRhGe2} emerges only at much lower temperatures ($T_\text{C}\approx 0.14$\,K) \cite{hirose2011magnetic}. Moreover, previous investigations of \ce{CeCoGe2} have been limited to arc-melted polycrystalline samples of relatively poor quality \cite{mun2004kondo}. These considerations motivated the present study in which we investigate the low-temperature physical properties of \ce{CeCoGe2} single crystals and search for signatures of superconductivity. 

This paper is organized as follows. First, we present and discuss the magnetic (section III A), thermodynamic (III B), and electrical transport properties (III C) of \ce{CeCoGe2} single crystals grown by the In-flux method and compare our results to those obtained previously on polycrystalline samples. Then, in section III D, we address the role of persistent intrinsic Co vacancies and describe our attempts to suppress them by controlling the starting stoichiometry and the temperature profile during crystal growth. We ascribe the absence of superconductivity in \ce{CeCoGe2} to strong disorder scattering of charge carriers induced by these vacancies.

\begin{figure*}[t!]
\centering
 \includegraphics[width=0.9\textwidth]{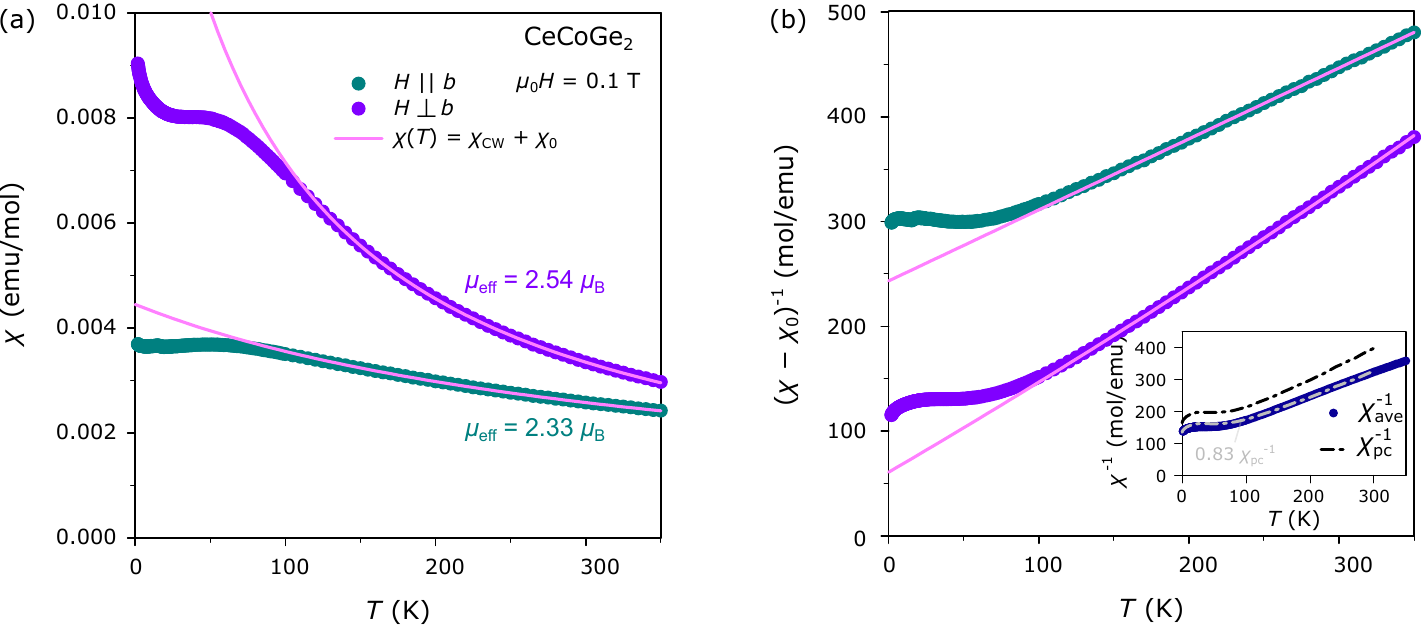}
\caption{Temperature-dependent (a) magnetic susceptibility and (b) inverse susceptibility of a plate-like \ce{CeCoGe2} single crystal as grown from indium flux. There exists magnetic anisotropy with $\chi$ being about two times larger perpendicular to the crystallographic $b$ axis compared to measurements with an applied field parallel to $b$. Modified Curie-Weiss fits of $\chi(T)$ at temperatures above $\approx 150$\,K yield effective magnetic moments of $2.54\,\mu_\text{B}$ and $2.33\,\mu_\text{B}$ for $H\perp b$ and $H \parallel b$, respectively. Inset in (b) shows the single crystal average $\chi_\text{ave}=\frac{2}{3}\chi_{\perp b}+\frac{1}{3}\chi_{\parallel b}$ compared to data for polycrystalline arc-melted samples by Mun \textit{et al.} \cite{mun2004kondo}. Scaling the polycrystalline data by a factor 0.83, which corresponds to the phase purity of polycrystalline samples synthesized in the course of this work following the same recipe as in Ref.\,\cite{mun2004kondo}, yields near-perfect agreement with $\chi_\text{ave}$.
}
\label{Fig3}
\end{figure*}

\section{Experimental Methods}
\subsection{Crystal growth}
\ce{CeCoGe2} single crystals were grown from In flux. Polycrystalline precursor ingots were first prepared by arc melting stoichiometric amounts of Ce (99.95\,\%), Co (99.95\,\%) and Ge (99.9999\,\%) under an inert Ar atmosphere. The resulting ingots were crushed and hand-ground into a coarse powder, which was subsequently mixed with In in a molar ratio of 1\,:\,30. The mixtures were loaded into alumina crucibles, sealed in evacuated quartz tubes, and heated to 1150\,$^\circ$C at a rate of 100\,$^\circ$C/h. After homogenization at this temperature for 48\,h, the samples were slowly cooled to 700\,$^\circ$C at a rate of 2\,$^\circ$C/h. The excess In flux was removed by centrifugation, and residual flux was eliminated by etching the crystals in diluted hydrochloric acid. The obtained crystals were plate-like, with thicknesses of up to a few hundred micrometers and lateral dimensions of up to a few square millimeters. The crystallographic orientation of the samples was determined using a Photonic Science Laue diffractometer, and the composition and microstructure were investigated in a scanning electron microscope equipped with a back-scattered electron detector and energy dispersive x-ray spectroscopy. The latter revealed homogeneous microstructures with uniform elemental distributions and no detectable impurity phases or residual flux.

\subsection{Characterization}
Single-crystal x-ray diffraction experiments were conducted at room temperature
using a Bruker D8 Venture diffractometer equipped with an Incotec I$\mu$S microfocus source using Mo K-$\alpha$ radiation ($\lambda = 0.71073$\,\AA). Data were collected using a PHOTON II CPAD detector and processed using Bruker SAINT software. The initial crystallographic model was obtained via intrinsic phasing methods in SHELXT. The refined structures were obtained via least-squares refinements using SHELXL2018. For as grown \ce{CeCoGe2} crystals, our refinements yield a unit cell volume of 300.12(11)\,\AA$^3$. After annealing at 600\,$^\circ$C for one week, we see a volume contraction of the unit cell down to 297--298\,\AA$^3$ for the annealed crystals. Specific heat and magnetic measurements were performed for the as-grown crystals, while electrical resistivity was measured for both as-grown and annealed crystals, revealing comparable residual resistivities and overall behavior but slightly modified low-temperature exponents as discussed in section III C.

Magnetic susceptibility ($\chi$) and magnetization ($M$) measurements were carried out in a Quantum Design (MPMS3) magnetometer using the VSM mode. Susceptibility was measured in a field of 1\,kOe applied parallel and perpendicular to a single crystal's $b$-axis.  Magnetization measurements at 1.8\,K were linear in field to 65\,kOe for both crystal orientations and showed the same anisotropy as $\chi$ (see Fig.\,7 in the Appendix).

Heat-capacity measurements were performed using the two-tau thermal relaxation method in a Quantum Design Physical Property Measurement System (PPMS). For measurements below 1.8\,K and down to 0.37\,K, a $^3$He refrigerator insert was employed. The electrical resistivity from 1.8 to 300\,K and up to 5\,T was also measured in a PPMS making use of a standard four-probe configuration with platinum wires spot-welded to the \ce{CeCoGe2} single crystals. Resistivity measurements down to 0.02\,K were performed in a Proteox dilution refrigerator.

\begin{figure*}[t!]
\centering
 \includegraphics[width=0.95\textwidth]{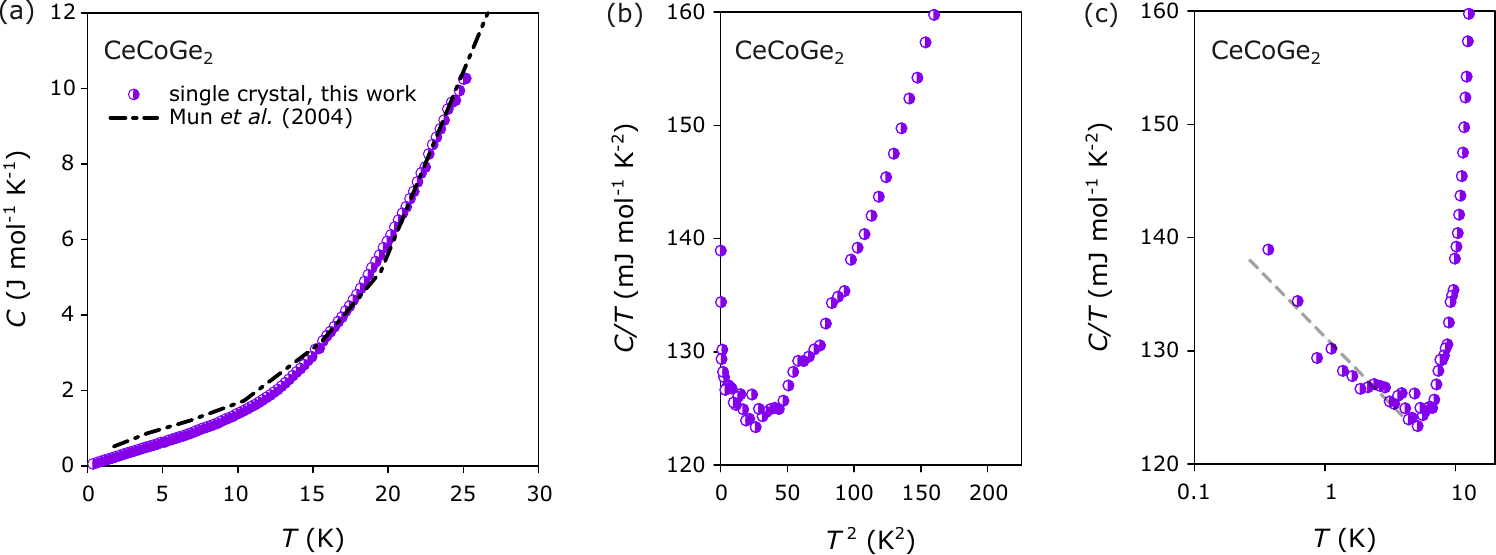}
\caption{Low-temperature specific heat of \ce{CeCoGe2}. (a) Comparison of as-grown \ce{CeCoGe2} single crystals from this work and arc-melted polycrystalline samples from Ref.\,\cite{mun2004kondo}. (b) The low-temperature slope of the specific heat $C/T$, measured down to 0.37\,K, reveals a heavy-fermion ground state with a  Sommerfeld coefficient of around 120\,mJ\,mol$^{-1}$\,K$^{-2}$ and an additional upturn of $C/T$ at the lowest temperatures, which, as shown in (c), roughly follows a logarithmic behavior.}
\label{Fig4}
\end{figure*} 

\section{Results and Discussion}
\subsection{Magnetic susceptibility}
Figure 3 shows the magnetic susceptibility [Fig.\,3(a)] and inverse susceptibility [Fig.\,3(b)] of an as-grown, plate-like \ce{CeCoGe2} single crystal, measured with magnetic fields ($\mu_0 H=0.1\,$T) applied perpendicular ($H\perp b$) and parallel ($H\parallel b$) to the crystallographic $b$ direction. It can be seen that there exists sizeable magnetic anisotropy,  with $\chi_{\perp b}$ being approximately two times larger than $\chi_{\parallel b}$, opposite to \ce{CeNiGe2}, \ce{CeRhGe2} and \ce{CePtSi2}, where $\chi_{\perp b}\ll \chi_{\parallel b}$ \cite{jung2002magnetocrystalline,hirose2011magnetic}. We fit the high-temperature data using a modified Curie-Weiss law, 
\begin{equation}
    \chi(T)=\chi_\text{CW}+\chi_0=C/(T-\theta)+\chi_0\,.
\end{equation}

\noindent Here, $\theta$ represents the paramagnetic Curie-Weiss temperature, $C=(N_\text{A}\mu_\text{eff}^2)/(3k_\text{B})$ and $\chi_0=\chi_\text{P}+\chi_\text{dia}$ is a constant term including Pauli-paramagnetic and diamagnetic contributions. From our fits, $\chi_0\approx 3-4\times10^{-4}\,$emu/mol for both field directions. Moreover, we determine effective moments of around $\mu_\text{eff}\approx 2.54 \,\mu_\text{B}$ for $H\perp b$, in excellent agreement to the value for the free Ce$^{3+}$ ion, and $\mu_\text{eff}\approx 2.33 \,\mu_\text{B}$ for $H\parallel b$, slightly smaller than the free-ion value (2.54\,$\mu_\text{B}$/Ce) with large negative Curie-Weiss temperatures, $\theta \approx -74$\,K and $-359$\,K for $H\perp b$ and $H\parallel b$, respectively. Since $\theta_{H\parallel b}$ exceeds the temperature range over which the fit was performed, there is likely a sizeable uncertainty in the fitted value of $\mu_\text{eff}$ for $H\parallel b$.

Below 100\,K, $\chi(T)$ develops a shallow bump, which may be attributed to crystal electric field (CEF) excitations, followed by an additional upturn at the lowest temperatures, likely due to magnetic impurities. Because a similar Curie tail at low temperatures has been observed in \ce{LaCoGe2}\cite{mun2004kondo}, it is possible that these features stem from small amounts of Co-based impurity phases or magnetic defects related to the Co sublattice. 

We also calculated the average susceptibility $\chi_\text{ave}=\frac{2}{3}\chi_{\perp b}+\frac{1}{3}\chi_{\perp b}$ obtained by averaging the single-crystal data and compared it to the polycrystalline data by Mun \textit{et al.} \cite{mun2004kondo} in the inset of Fig.\,3(b). Since $a=4.25\,$\AA$\,\approx c=4.20\,$\AA$\,\ll b=16.74\,$\AA, no significant anisotropy is expected in the $ac$ plane -- the lack of which was also confirmed for \ce{CeNiGe2}, \ce{CeRhGe2} and \ce{CePtSi2}. We find that $\chi_\text{ave}$ is in good agreement with $\chi_\text{pc}$ from Ref.\,\cite{mun2004kondo}, especially after scaling by a constant factor of 0.83 [grey line in Fig.\,3(b)], which roughly corresponds to the phase purity ($80-90\,$\%) of polycrystalline samples synthesized and investigated in the course of this work following the exact same recipe as in Ref.\,\cite{mun2004kondo}. 

\subsection{Specific heat}
In Fig.\,4, we show the low-temperature specific heat $C(T)$ of \ce{CeCoGe2}. Above 2\,K, our data are in excellent agreement with previously reported results on arc-melted polycrystalline samples by Mun \textit{et al.}\,\cite{mun2004kondo} [see Fig.\,4(a)], exhibiting a pronounced linear term below 10\,K that corresponds to a Sommerfeld coefficient of $\gamma \approx 120$\,mJ\,mol$^{-1}$\,K$^{-2}$, consistent with a heavy-fermion ground state as predicted by the phase diagram in Fig.\,2. At the lowest temperatures, an additional upturn of $C/T$ with decreasing temperature is observed but no signs of ordering are apparent down to 0.37\,K [Fig\,4(b)], consistent with the electrical resistivity discussed in the next sections, which shows no signs of a phase transition down to 20\,mK. A similar low-temperature upturn in $C/T$ has previously been reported for off-stoichiometric \ce{CeCo_{0.89}Ge2} \cite{pecharsky1991ceco}. Notably, the low-temperature anomaly exhibits little dependence on applied magnetic field (see Ref.\,\cite{pecharsky1991ceco}), suggesting that it may be an intrinsic feature rather than a Schottky anomaly arising from magnetic point defects or other extrinsic mechanisms. As seen in Fig. 4(c), the upturn approximately follows a logarithmic temperature dependence, $C/T \sim -\ln (T/T_0)$, a behavior often observed in the vicinity of quantum critical points, where $k_\text{B}T_0$ represents a characteristic energy scale associated with the quantum critical fluctuations \cite{lohneysen2007fermi}. In the present case, however, the magnitude of the increase is relatively small ($\approx15-20\,\%$), which may indicate that \ce{CeCoGe2} is located near, but not exactly at, the putative quantum critical point in the phase diagram. We obtain $T_0= 2-5\,$K (depending on the temperature range of the fit), consistent with this interpretation.

\begin{figure*}[t!]
\centering
 \includegraphics[width=0.9\textwidth]{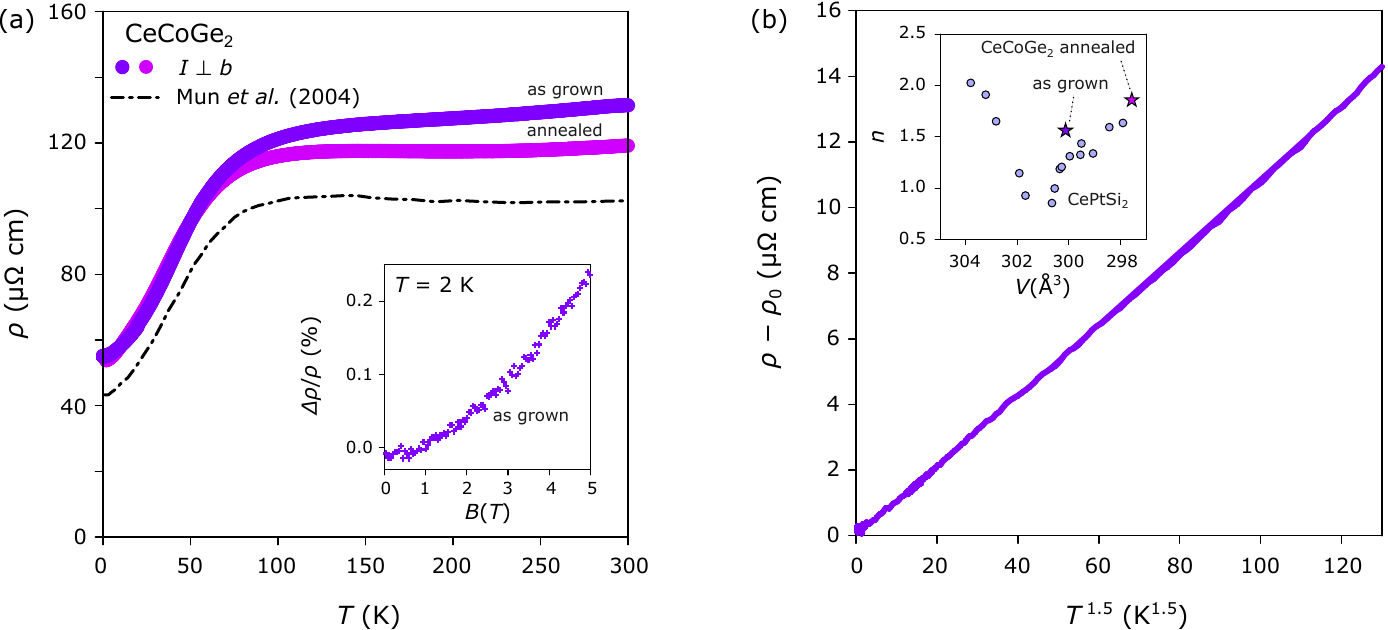}
\caption{Temperature-dependent electrical resistivity of \ce{CeCoGe2}. (a) shows a comparison between our single crystals, where current was applied perpendicular to $b$, and the results for arc-melted polycrystalline \ce{CeCoGe2} by Mun \textit{et al.} \cite{mun2004kondo} (black dashed-dotted line), which is about a factor of 1.4 smaller, while showing a comparable residual resistivity ratio RRR $\approx2.5$. Inset shows negligible field dependence of the magnetoresistance at $T=2\,$K. (b) Low-temperature scaling of $\rho(T)$ follows a non-Fermi-liquid (NFL) behavior with $\rho(T)=\rho_0 + AT^{n}$ ($n\approx 1.5$). Inset shows unit-cell volume dependence of the resistivity exponent $n$ for hydrostatic-pressure-tuned \ce{CePtSi2} \cite{nakano2009pressure} and \ce{CeCoGe2} at ambient pressure (this work).
}
\label{Fig5}
\end{figure*} 

\subsection{Electrical resistivity}
Figure 5 shows the temperature-dependent electrical resistivity $\rho(T)$ of an as-grown \ce{CeCoGe2} single crystal. Owing to the thin plate-like shape of the samples, measurements were only performed with current along the crystallographic $ac$ plane. Different crystals from the same batch yield reproducible results with comparable values for the residual resistivities ($\rho_0\approx 50-60\,\mu\Omega$\,cm) and residual resistivity ratios (RRR $=2.3-2.5$). Both $\rho_0$  and RRR are also comparable to those reported for polycrystalline samples from Ref.\,\cite{mun2004kondo}, indicating that grain boundary scattering is not dominating the electrical transport. Indeed, as will be discussed later, the rather low RRR and sizeable $\rho_0$ is instead related to intrinsic Co vacancies, which lead to strong random potential fluctuations and impurity scattering of charge carriers. In contrast to \ce{CePtSi2}, there is almost no field dependence of $\rho(T)$ (less than 0.3\,\% at 5\,T), with a vanishingly small and positive transverse magnetoresistance [see inset of Fig.\,5(a)].

We find that the low-temperature $\rho(T)$ behavior of \ce{CeCoGe2} does not follow the conventional quadratic temperature dependence of Fermi liquids $\rho_\text{FL}=\rho_0+AT^2$, but can instead be well described by a power law $\rho(T)=\rho_0+AT^{n}$ with $n\approx 1.5$ [see Fig.\,5(b)]. This behavior persists over an extended temperature range up to $T\approx 25\,$K and appears to be correlated with the unit-cell volume (inset Fig.\,5) as annealed crystals with contracted $V$ display exponents closer to $n\approx 1.8$. It is known that disorder can modify the low-temperature exponent of $\rho(T)$, and anomalous non-Fermi-liquid (NFL) exponents, $n=3/2$, have been associated with glassy behavior \cite{miranda2005disorder} and observed in various spin-glass and mictomagnetic systems \cite{mydosh2011colloquium}. Another possibility is that the NFL exponent in \ce{CeCoGe2} arises from its proximity to the putative quantum critical point indicated by the temperature--pressure phase diagram in Fig.\,2 \cite{hertz2018quantum,millis1993effect,moriya1995anomalous,kambe1996application,rosch1999interplay}. For comparison, \ce{CePtSi2} has also revealed NFL behavior under pressure, close to the QCP and its superconducting dome. The inset of Figure 5(b) shows the exponent $n$ of \ce{CePtSi2} from Ref.\cite{nakano2009pressure} as a function of the unit-cell volume. At ambient pressure, $\rho(T)$ shows Fermi-liquid behavior with $n=2$, but near the critical volume, the resistivity exponent drops to $n\approx 1$, and then increases to $n\approx 1.5-1.6$ at higher pressures (\textit{i.e.} smaller volumes $V\approx 298-300\,$\AA$^3$). This may imply that \ce{CeCoGe2} resides on the paramagnetic side of the quantum phase transition, in close proximity to, but not exactly at, the QCP. However, follow-up studies on higher-quality crystals and with continuous tuning parameters, such as magnetic field or uniaxial strain are required to determine whether \ce{CeCoGe2} can be tuned to quantum criticality. No signatures of superconductivity or magnetic order are observed down to 20\,mK (Fig.\,6).

\subsection{Intrinsic defects}
To elucidate the reason for the rather low RRR and high $\rho_0$, we conducted single-crystal x-ray diffraction (XRD) experiments on several \ce{CeCoGe2} single crystals. Our results show variation in the Co occupancy, consistent with previous reports on \ce{RECo_xGe2} germanides, which show a range of Co from 0.34 to 0.96 depending on the rare-earth (RE) element \cite{meot1985nouveaux,francois1990nouveaux}.
Refinements of single-crystal XRD patterns reveal Co vacancies of the order of 4\,\% ($x=0.96$) in the crystals, which were grown from nominally stoichiometric \ce{CeCoGe2} precursors dissolved in In flux, along with a slight deficiency of Ge on the Ge2 site ($\approx98.7(4)$\,\% occupancy). This prompted us to optimize the stoichiometry and try to suppress the Co vacancies by adding additional Co and Ge to the arc-melted precursor material used during the flux growth, resulting in nominal compositions of \ce{CeCo_{1+2x}Ge_{2+x}}.

The observed phase of the resulting crystals and the concentration of Co vacancies as well as the residual resistivity ratios and resistivity exponents are summarized in Table 1. For $x>0.2$, we find that \ce{CeCo2Ge2} crystals are formed, which belong to the tetragonal \ce{ThCr2Si2} structure type. Interestingly, against our expectations, for $x<0.2$, the obtained \ce{CeCoGe2} crystals display an even larger number of Co vacancies and even lower RRR values. We attribute this to a competition between the 112 (\ce{CeCoGe2}) and 122 (\ce{CeCo2Ge2}) phases. Even for nominally stoichiometric \ce{CeCoGe2}, arc-melted samples displayed about $10-20$\,\% of the 122 phase, even after annealing for three weeks at 900\,$^\circ$C. We conclude that this behavior suggests either (i) the 112 phase is intrinsically unstable in its stoichiometric form and requires a small number of vacancies for stabilization, or (ii) a strong competition exists between the 112 and 122 phases, leading to off-stoichiometric variants of \ce{CeCoGe2} as \ce{CeCo2Ge2} forms and depletes the molten solution of Co.

\begin{table}[b!]
\centering
\caption{Influence of starting stoichiometry on phase formation and transport quality of Ce-Co-Ge crystals grown by flux. Deviations from the nominal \ce{CeCoGe2} composition do not improve crystal quality; instead, they increase Co deficiency and reduce the RRR, or stabilize the competing tetragonal \ce{CeCo2Ge2} phase. The exponent $n$ is obtained from power law fits $\rho(T)=\rho_0+AT^n$  to the low-temperature resistivity over an extended range of temperatures $T\lesssim 25\,$K.}

\label{tab:3x5}
\begin{tabular}{ccccc}
\hline
composition \,\,& observed phase \,\,& Co vacancies \,\,& RRR & $n$  \\ \hline

\ce{CeCoGe2} & \ce{CeCoGe2} & 4\,\%  & 2.5  & 1.5--1.8 \\

\ce{CeCo_{1.2}Ge_{2.1}} & \ce{CeCoGe2} & 11\,\%  & 1.5  & 1.3--1.6 \\

\ce{CeCo_{1.4}Ge_{2.2}}, &
\multirow[c]{2}{*}{\ce{CeCo2Ge2}} &
\multirow[c]{2}{*}{--} &
\multirow[c]{2}{*}{5.3} &
\multirow[c]{2}{*}{2} \\

\ce{CeCo_{1.6}Ge_{2.3}} & & & & \\

\hline
\end{tabular}
\end{table}

The fact that 4\,\% Co vacancies result in a serious deterioration of the RRR and a sizeable residual resistivity, $\rho_0\approx 50-60\,\mu \Omega$\,cm, suggests that Co $3d$ bands contribute significantly to the density of states near the Fermi level $E_\text{F}$ (in agreement with density functional theory calculations \cite{choi2013observation}) and that Co site vacancies introduce strong disorder and decoherence for the Co $3d$ bands. Another possibility is that Co vacancies introduce nonhybridizing impurity states near $E_\text{F}$ that can strongly scatter charge carriers from the bulk conduction bands, which is not uncommon for transition-metal site substitutions \cite{krishnan2020unconventional,garmroudi2022anderson,reumann2022thermoelectric,garmroudi2023pivotal,jha2024unexpected,parzer2025mapping,parzer2024semiconducting,rogl2022understanding}. Figure 6 compares the temperature-dependent resistivity, normalized to its room-temperature value $\rho(T)/\rho(300\,$K), of \ce{CeCoGe2} (this work) with single-crystal data for \ce{CePtSi2} \cite{hirose2011magnetic,nakano2013transport}, \ce{CeNiGe2} \cite{jung2002magnetocrystalline} and \ce{CeRhGe2} \cite{hirose2011magnetic} from the literature. It is apparent that other \ce{Ce$TX_2$} systems also suffer from rather low residual resistivity ratios, which can probably be related to intrinsic transition-metal vacancies as well. Indeed, intrinsic vacancies are ubiquitous in materials crystallizing in the \ce{CeNiSi2} and the vacancy concentration depends sensitively on the rare-earth and transition metal element \cite{meot1985nouveaux,francois1990nouveaux}. In the case of \ce{CePtSi2}, however, $\rho_0$ decreases significantly under applied magnetic field \cite{nakano2013transport}, indicating that the low RRR may be dominated by strong magnetic scattering and spin fluctuations near the antiferromagnetic transition, in contrast to the field-independent behavior observed in \ce{CeCoGe2} (see inset Fig.\,5a).

\begin{figure}
\centering
 \includegraphics[width=0.42\textwidth]{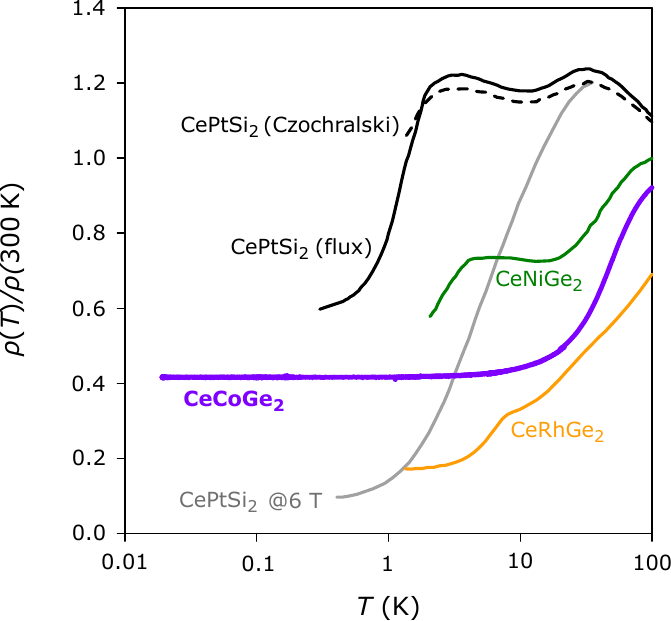}
\caption{Comparison of temperature-dependent resistivity normalized to room temperature $\rho(T)/\rho(300\,\text{K})$ for various single-crystalline \ce{Ce$TX_2$} systems. Next to \ce{CeCoGe2}, other \ce{Ce$TX_2$}-based compounds also suffer from poor residual resistivity ratios. This suggests that intrinsic defects play an important role in the low-temperature transport properties of the \ce{CeNiSi2} materials family in general.
}
\label{Fig2}
\end{figure} 

\section{Conclusion}
We compared the chemical and hydrostatic pressure dependence of locally noncentrosymmetric \ce{Ce$TX$2}-based materials, crystallizing in a pseudotetragonal, orthorhombic ($Cmcm$) structure, where $T$ is a transition metal from the Fe, Co, Ni and Cu groups and $X$ is either Si or Ge. The period of the transition metal and the unit-cell volume sensitively control the hybridization of the Ce 4$f$ states with the conduction electrons, resulting in a putative QCP at a critical volume $V_\text{c}\approx 300\,${\AA}$^3$, from which unconventional superconductivity emerges in \ce{CePtSi2} and \ce{CeRhGe2}, both of which require hydrostatic pressure tuning. The large upper critical fields in \ce{CePtSi2}, about an order of magnitude above the Pauli limiting field, suggest a possible spin-triplet pairing. We identify \ce{CeCoGe2} as a promising heavy-fermion compound sitting in proximity of the putative quantum critical point at ambient pressure. However, no superconductivity is detected down to 20\,mK, which we attribute to the rather large residual resistivity. Single-crystal x-ray diffraction reveals intrinsic disorder in the form of Co-site vacancies, even in nominally stoichiometric samples that strongly affect electronic transport. Composition tuning further shows that these defects are likely intimately linked to a competing tetragonal phase (\ce{CeCo2Ge2}) that forms at comparable temperatures.

\section{Acknowledgements}
Work at Los Alamos National Laboratory was performed under the auspices of the U.\,S. Department of Energy, Office of Basic Energy Sciences, Division of Materials Science and Engineering. F.\,G. acknowledges a
Director’s Postdoctoral Fellowship through the
Laboratory and Directed Research \& Development (LDRD) program. S.\,M.\,T. acknowledges support from the Los Alamos LDRD program.
Scanning electron microscopy and energy-dispersive x-ray measurements were performed at the Center for Integrated Nanotechnologies, an Office of Science User Facility operated for the U.S. Department of Energy (DOE) Office of Science.

\section{Appendix}
Figure 7 shows the field-dependent magnetization of an as-grown \ce{CeCoGe2} single crystal. Magnetization data were obtained at 1.8\,K with fields applied perpendicular and parallel to the crystallographic $b$-axis, showing linear behavior over the entire measured field range with no signs of saturation reaching values up to $\approx 0.09$ and $0.036\,\mu_\text{B}$/Ce at 6.5\,T for $H\perp b$ and $H\parallel b$, respectively.

\begin{figure}
\centering
 \includegraphics[width=0.42\textwidth]{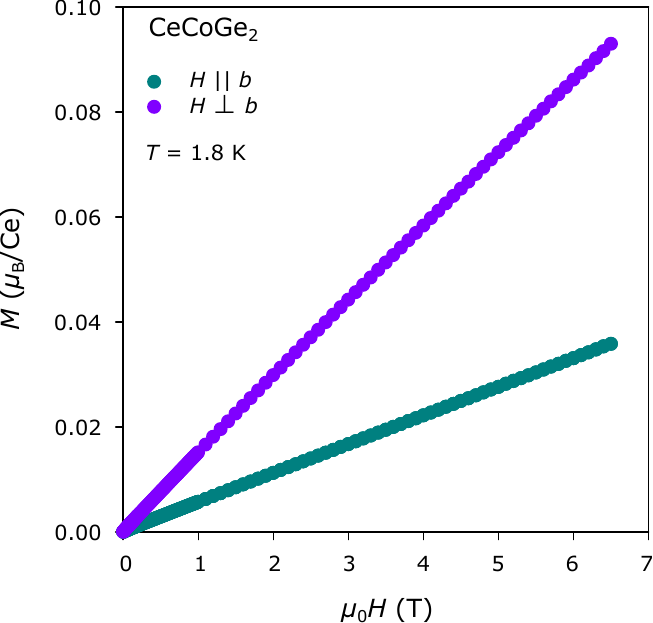}
\caption{Field-dependent magnetization of \ce{CeCoGe2} single crystal as grown from In flux. Data were obtained at 1.8\,K with fields applied perpendicular and parallel to the crystallographic $b$-axis, showing linear behavior over the entire measured field range.
}
\label{Fig2}
\end{figure} 

\clearpage
%

\end{document}